\documentclass[12pt]{article}
\usepackage{amsmath,amssymb,amscd,cite}
\newtheorem{thm}{Theorem}[section]
\newtheorem{prop}[thm]{Proposition}
\newtheorem{lem}[thm]{Lemma}
\newtheorem{cor}[thm]{Corollary}
\newtheorem{defi}[thm]{Definition}
\newcommand{\pf}{{\bf Proof. \ }}
\newcommand{\qed}{\hfill $\Box$ \\}

\font\msbm=msbm10 at 12pt
\newcommand{\Z}{\mbox{\msbm Z}}
\newcommand{\ZZ}{\mbox{\msbm Z}}

\newcommand{\FF}{\mbox{\msbm F}}
\newcommand{\F}{\mbox{\msbm F}}
\newtheorem{rem}[thm]{Remark}
\newtheorem{ex}[thm]{Example}

\newcommand{\ord}{ord}

\begin{document}
\author{ Kenza Guenda and T. Aaron Gulliver
\thanks{ T. Aaron Gulliver is
with the Department of Electrical and Computer Engineering,
University of Victoria, PO Box 3055, STN CSC, Victoria, BC, Canada
V8W 3P6. email: agullive@ece.uvic.ca.}}

\title{Construction of Cyclic Codes over $\mathbb{F}_2+u\mathbb{F}_2$ for DNA Computing}
\date{}

\maketitle

\begin{abstract}
We construct codes over the ring $\mathbb{F}_2+u\mathbb{F}_2$ with
$u^2=0$. These code are designed for use in DNA computing
applications. The codes obtained satisfy the reverse complement
constraint, the $GC$ content constraint and avoid the secondary
structure. they are derived from the cyclic complement reversible
codes over the ring $\mathbb{F}_2+u\mathbb{F}_2$. We also construct
an infinite family of BCH DNA codes.
\end{abstract}

Deoxyribonucleic acid (DNA) contains the genetic program for the
biological development of life. DNA is formed by strands linked
together and twisted in the shape of a double helix. Each strand is
a sequence of four possible nucleotides, two purines; adenine $(A)$,
guanine $(G)$ and two pyrimidines; Thymine $(T)$ and cytosine $(C)$.
The ends of a DNA strand are chemically polar with $5'$ and $3'$
ends, which implies that the strands are oriented. Hybridization,
known as base pairing, occurs when a strand binds to another strand,
forming a double strand of DNA.

The strands are linked following the Watson-Crick model. Every $(A)$
is linked with a $(T)$, and every $(C)$ with a $(G)$, and vice
versa. We denote the complement of $X$ as $\hat{X}$, i.e.,
$\hat{A}=T,\hat{T}=A,\hat{G}=C$ and $\hat{C}=G$. The pairing is done
in the opposite direction and the reverse order. For instance, the
Watson-Crick complementary (WCC) strand of $3'-ACTTAGA-5'$ is the
strand $5'-TCTAAGT-3'$. Non-specific hybridization occurs when
hybridization between a DNA strand and its Watson-Crick complement
does not take place, or when a DNA strand hybridizes with the
reverse of a distinct strand. Another non-specific hybridization is
when a strand folds back onto itself, forming a so-called
``secondary structure''.

DNA computing is the fusion of the world of genetic data analysis
and the science of computation in order to tackle computationally
difficult problems. This new area was born in 1994 when
Adleman~\cite{adleman} solved an instance of a hard (NP-complete)
computational problem, namely the directed traveling Salesman
problem on a graph with seven nodes. Their approach was based on the
WCC property of DNA strands. Since then, numerous studies have built
on their research and expanded DNA computing to solve other
mathematical problems~\cite{adleman2,boneh,lipton}. Furthermore,
since there are $4^n$ possibly single DNA strands of length $n$
which can be quickly and cheaply synthesized, Mansuripur et
al.~\cite{mansuripur} showed that DNA codewords can be used as ultra
high density storage media. Other application make use of the DNA
hybridization phenomena~\cite{shoemaker}.

A block code is called a DNA code if it satisfies some of the
following constraints:
\begin{enumerate}
\item the Hamming constraint for a distance $d$,
\item the reverse-complement constraint,
\item the reverse constraint, and
\item the fixed $GC$-constraint.
\end{enumerate}
The purpose of the first three constraints avoid non-desirable
hybridization between different strands. The fixed $GC$-constraint
ensures all codewords have similar thermodynamic characteristics,
which allows parallel operations on DNA sequences. Milenkovic and
Kashyap~in~\cite{olgica} proved that when designing a DNA code a
fifth constraint should be added in order to make secondary
structure less likely to happen. Secondary structure causes
codewords to become computationally inactive, as the codewords have
low chemical activity. This defeats the read-back mechanism in a DNA
storage system by $30 \%$ as reported by Mansuripur et
al.~\cite{mansuripur}. Milenkovic and Kashyap~\cite{olgica} used the
Nussinov-Jacobson algorithm~\cite{nussinov} to prove that the
presence of a cyclic structure reduces the complexity of testing DNA
codes for secondary structure, and also simplifies DNA sequence
fabrication. Another advantage of the design of cyclic codes, as
pointed out by Siap et al.~\cite{siap}, is that the complexity of
the dynamic programming algorithm to find the largest common
subsequence between any two codewords in a cyclic code will be less
than that of any other codes. there have been numerous papers on the
design of DNA codes~\cite{abualrub,abualrub1,gaborit,siap}. Gaborit
and King~\cite{gaborit} and Abualrub et al~\cite{abualrub}
constructed DNA codes over $GF(4)$. Siap et al.~\cite{siap}
constructed cyclic DNA codes considering the $GC$-content constraint
over $\mathbb{F}_2[u]/(u^2-1)=\{0,1,u,u+1\}$, where $u^2=1$, and
used the deletion distance.

In this paper, we construct cyclic linear codes suitable for
DNA-computing. They are derived from cyclic reverse-complement codes
over the ring $R=\mathbb{F}_2+u\mathbb{F}_2$, where $u^2=0$. We give
infinite families of DNA codes with either fixed $GC-$content, or
with few weights in order to obtain DNA codes with a large fixed
$GC-$content after removing codewords that violate the $GC-$content
constraint. Since our codes are cyclic, this can  be done easily as
noted by Abualrub et al.~\cite{abualrub1}. Furthermore, we will
benefit from the fact that this ring contains $\mathbb{F}_2$ as a
subring and has properties in common with $\mathbb{Z}_4$.
In addition, techniques for implementation and decoding have been
developed~\cite{udaya1}. These codes can also correct certain burst
errors. We also construct BCH codes over this ring, and BCH DNA
codes. BCH codes over fields are well known, hence we translate the
properties of BCH codes over $\FF_2$ to $R$. Previously~
Shankar~\cite{shankar} constructed BCH codes over the rings $\ZZ_m$.
Calderbank and Sloane~\cite{CS} gave BCH codes over $\ZZ_{p^a}$ as a
Hensel lift of BCH codes from fields to rings. We construct BCH
codes over the ring $R$ without using a lift. Furthermore, decoding
algorithms exist such as that given by Bonnecaze and
Udaya~\cite{udaya2}. For the reasons given above, these codes are
very appropriate for DNA computing.

\section{Preliminaries}

The ring considered here is the ring $R=\mathbb{F}_2+u\mathbb{F}_2$,
where $u^2=0$. A linear code over this ring is a module over $R$.
Codes over this ring were introduced by Bachoc~\cite{bachoc} and
studied by Bonnecaze and Udaya~\cite{udaya1, udaya2}, Dougherty et
al.~\cite{type2,dougherty}, Gulliver and
Harada~\cite{gulliver1,gulliver2}, and more recently by Abualrub and
Siap~\cite{abualrub}.

The ring $R$ contains four elements $\{0,1,u,1+u\}$. This is a local
commutative ring with characteristic 2 and unique maximal ideal
$\langle u \rangle$. It is also a finite chain ring. It contains
unique chain ideals $0\subset \langle u \rangle \subset R$. The
field $\F_2$ can be seen as a subring  of $R$. This is an
interesting fact which will be useful later.

For linear codes over a chain ring, the rank of $\mathcal{C}$
denoted $rank(\mathcal{C})$ is defined as the minimum number of
generator of $\mathcal{C}$. In this paper, we only consider codes
with odd length. We define the Hamming weight of a codeword $x$ in
$\mathcal{C}$ as $w_H(x)=n_1(x)+n_u(x)+n_{u+1}(x)$, the Lee weight
of $x$ as $w_L(x)=n_1(x)+2n_u(x)+n_{u+1}(x)$, and the Euclidean
weight as $w_E(x)=n_1(x)+4n_u(x)+n_{u+1}(x)$. The Hamming, Lee and
Euclidean distances $d_H (\mathsf{x},\mathsf{y})$, $d_L
(\mathsf{x},\mathsf{y})$, $d_E (\mathsf{x},\mathsf{y})$ between two
vectors $\mathsf{x}$ and $\mathsf{y}$ are
$wt_H(\mathsf{x}-\mathsf{y})$, $wt_L(\mathsf{x}-\mathsf{y})$ and
$wt_E(\mathsf{x}-\mathsf{y})$, respectively. The minimum Hamming,
Lee and Euclidean weights, $d_H$, $d_L$ and $d_E$ of $C$ are the
smallest Hamming, Lee and Euclidean weights among all nonzero
codewords of $C$.

The elements $\{0,u,u+1,1\}$ of $R$ are in one to one correspondence
with the nucleotide DNA bases, $A, T, C, G $, such that
$0\rightarrow A$, $u\rightarrow T$, $u+1 \rightarrow C$ and $ 1
\rightarrow G$. We remark that for all $x\in R$, we have
\begin{equation}
\label{eq:hat}
 x+\hat{x}=u.
\end{equation}
We define the reverse of $x=x_0u_1\cdots x_{n-1}$ to be
$x^r=x_{n-1}x_{n-2}\cdots x_{1}x_0$. The complement of the codeword
$x=x_0x_1\cdots x_{n-1}$ is the vector $x^c=\hat{x_0}\hat{x_1}\cdots
\hat{x_{n-1}}$, and the reverse complement (also called the
Watson-Crick complement) is
$x^{rc}=\hat{x_{n-1}}\hat{x_{n-2}}\cdots \hat{x_{1}} \hat{x_{0}}$.

A linear code $\mathcal{C}$ over $R$ is said to be \textbf{cyclic}
if it is invariant under a cyclic shift, i.e.,
$(x_{n-1},x_0,\ldots,x_{n-2})\in \mathcal{C}$ provided the codeword
$(x_0,x_1,\ldots,x_{n-2},x_{n-1})$ is in $\mathcal{C}$. A code
$\mathcal{C}$ is said to satisfy \textbf{the reverse constraint} if
$H(x^r,y)\ge d$ for all $x,y \in \mathcal{C}$, including $x=y$. A
code $\mathcal{C}$ is said to satisfy \textbf{the Hamming
constraint} if for any two different codewords $x, y \in
\mathcal{C}$, $H(x,y) \geq d$. A code $\mathcal{C}$ is said to
satisfy \textbf{the reverse-complement constraint} if for any two
codewords $x,y \in \mathcal{C}$ (where $x$ might equal $y$),
$H(x^{rc},y) \geq d$. A code $\mathcal{C}$ is said to satisfy
\textbf{the fixed $GC-$content constraint} if any codeword $x\in
\mathcal{C}$ contains the same number of $G$ and $C$ elements. A
code is called a DNA code if it satisfies some or all of the
conditions above.

\section{Cyclic Codes over $R$}

In this section, we consider the cyclic codes of $R$ since our goal
is the study of cyclic DNA codes. The results of the reference above
are reviewed and extended. We also introduce the concept of $BCH$
codes over $R$. Only codes of odd length $n$ are examined.

The cyclic codes of odd length $n$ over $R$ are principal ideals of
the ring $R_n=\frac{R[x]}{\langle x^n-1 \rangle}$. Hence knowing the
factorization of $x^n-1$ is important.
\begin{lem}
\label{lem:sur1} (\cite[Theorem 3.3]{survey}) Let $R$ be a finite
chain ring with residual field $K$ of characteristic $p$. Let $n$ be
an integer such that $(n,p)=1$, hence $x^n-1$ factors uniquely as
basic irreducible polynomials. Furthermore, there is a one to one
correspondence between the factors of $x^n-1$ over $R$ and the
factors of $x^n-1$ over $K$.
\end{lem}
From Lemma~\ref{lem:sur1}, we have a one to one correspondence
between the factors of $x^n-1$ in $R$ and the factors of $x^n-1$ in
$\F_2$. However, since $F_2 \subset R$, the factors of $x^n-1$ in
$R$ are the same as in $\F_2$. This gives the following Lemma.
\begin{lem}
\label{lem:sur} If $n$ is odd then the factorization of $x^n-1$ into
irreducible polynomials over $R$ is the same as the factorization
over $F_2$.
\end{lem}

\begin{thm}
\label{th:gen} Let $\mathcal{C}$ be a cyclic code over $R$. Hence
$R_n$ is a principal ideal ring and there exist unique pairwise
coprime polynomials $F_0,F_1,F_2$ in $\F_2[x]$ such that $F_0
F_1F_2=x^n-1$, and
\begin{equation}
\label{eq:gen} \mathcal{C}=\langle F_0F_2|u F_0\rangle = \langle
F_0F_2+uF_0\rangle.
\end{equation}
Moreover
\begin{equation}
\label{car:cons} |\mathcal{C}|=(2)^{2\deg F_{1}+\deg F_2},
\end{equation}
and
\begin{equation}
\label{eq:rank} rank(\mathcal{C})=\deg  F_1+ \deg F_2.
\end{equation}
\end{thm}
\pf The proof follows from Lemma~\ref{lem:sur} and~\cite[Theorems
3.4 and 3.5]{permounth}. \qed

From now on, for simplicity of notation, we will write the cyclic
code given in (\ref{eq:gen}) as
 \begin{equation}
\label{eq:gen}
 \mathcal{C}=\langle f_0|uf_1 \rangle= \langle
 f_0+uf_1\rangle,
\end{equation}
  such that $f_1 |f_0 |x^n-1$.
It is clear $f_0=F_0F_2$ and $F_0=f_1$. Hence from~(\ref{eq:rank}),
the rank of $\mathcal{C}$ is equal to
 \begin{equation}
 \label{eq:rnc}
 r=rank(\mathcal{C})=n-\deg f_1.
 \end{equation}
There are two binary cyclic codes associated with a cyclic code
$\mathcal{C}$ over $R$; the binary code $Res(\mathcal{C})=\{x\in
\mathbb{F}_2 | \exists y \in \mathbb{F}_2^n, x+uy \in \mathcal{C}\}$
and $Tor(\mathcal{C})= \{x\in \mathbb{F}_2 | ux \in \mathcal{C}\}$,
called respectively the residue code and the torsion code. It has
been proven that~\cite[p 2150]{udaya2} $Res(\mathcal{C})= \langle
f_0 \rangle$ and $Tor(\mathcal{C})=\langle f_1 \rangle$.

Now we will consider the minimum distance of codes over $R$. First
we prove the following Lemma.
\begin{lem}
\label{lem:distance} If $\mathcal{C}$ is a code over $R$, then
$d_L(\mathcal{C}) \leq 2d_H(C)$, and $d_E(\mathcal{C}) \le
4d_H(\mathcal{C})$
\end{lem}
\pf Given a vector with Hamming weight $d$, the highest possible Lee
weight is obtained if all the non-zero coordinates are $u$, in which
case it has Lee weight $2d_H$. The same applies for the Euclidean
weight except that this vector has Euclidean weight $4d_H$.
 \qed
\begin{thm}
\label{th:dis} Let $\mathcal{C}=\langle f_0|uf_1 \rangle$ be a
cyclic code over $R$ of odd length $n$. Then the minimum distance of
$C$ satsifies the following
\begin{itemize}
\item[(i)] $d_H(\mathcal{C})=d_H(Tor(\mathcal{C}))=d_H(\langle f_1 \rangle)$,
\item[(ii)] $d_L(\mathcal{C})\le \min (d_H(\langle f_0 \rangle ,2d_H(\langle f_1 \rangle))$,
\item[(iii)] $d_H(\mathcal{C})\leq \deg f_1+1$,
\item[(iv)] $\lfloor \frac{d_L -1}{2} \rfloor \leq \deg f_1 +1,$
\item[(v)] $\lfloor \frac{d_E -1}{4} \rfloor \leq \deg f_1+1.$
\end{itemize}
\end{thm}
\pf From~\cite[Theorem 4.2]{Ana}, we have that
$d_H(\mathcal{C})=d_H(Tor(\mathcal{C}))$. Part (ii) comes from the
fact that the codes $Tor(\mathcal{C})$ and $Res(\mathcal{C})$ are
binary cyclic codes generated by $f_1$ and $ f_0$, respectively, and
satisfy $u \langle f_1 \rangle \subset \mathcal{C}$ and $\langle f_0
\rangle \subset \mathcal{C}$. The dimension of $Tor(\mathcal{C})$ is
$n-\deg(f_1)$. By the Singleton bound we have
$d_H(Tor(\mathcal{C}))\leq \deg(f_1)+1$. Hence Part (iii) follows
from Part (i). Parts (iv) and (v) follow from Part (iii) and
Lemma~\ref{lem:distance}. \qed

\subsection{BCH Codes over $R$}

A BCH code of length $n$ and designed distance $\delta$ over a field
$\FF_q$, denoted by $BCH(n,\delta)_q$ is defined as a cyclic code
generated by $\mbox{lcm} (M_1,M_2,\ldots, M_{\delta-1})$, where the
$M_i$ are the minimal polynomial factors of $x^n-1$ over $\FF_q$.
The definition of BCH codes over $\FF_2$ can be  extended to the
ring $R=\F_2+u\F_2$, $u^2=0$. This follows from Lemma~\ref{lem:sur}
if $x^n-1=\prod_{i=0}^r M_i$ is the unique factorization of the
polynomial $x^n-1$ over $R$. The $M_i$ are minimal polynomial over
$\F_2$, each of which corresponds to a cyclotomic class modulo $n$.

\begin{defi}
Let $n, \delta_0, \delta_1$ be positive integers such that $1 \leq
\delta_1 \leq \delta_0 \leq n-1$. We define the $BCH$ code of length
$n$ and designed distance $(\delta_0,\delta_1)$ over $R$ to be the
cyclic code $\langle g_{\delta_0}, ug_{\delta_1} \rangle$, with
$g_{\delta_j}=\mbox{lcm}(M_i), 1\leq i \leq \delta_j-1$ where $0\leq
j \leq 1$ and $\delta_1 \leq \delta_0$. We denote this code by
$BCH(n,\delta_0,\delta_1)$.
\end{defi}
We have the following results concerning the rank and minimum
distance of the $BCH$ codes over $R$.
\begin{thm}
Let $\mathcal{C}$ be a $BCH(n,\delta_0, \delta_1)$ be a BCH code
over $R$ of length $n$ and designed distance $(\delta_0, \delta_1)$.
Then the following holds
\begin{itemize}
\item[(i)]  $\min (\delta_0, 2\delta_1) \le d_L(\mathcal{C}) \le \min(d_H(BCH(n,\delta_0)), 2d_H(BCH(n,\delta_1)))$
\item[(ii)] $\delta_i, 0 \le i\le 1$ can be assumed to be odd
\item[(iii)] If $\delta _1= 2w+1$, hence $rank(\mathcal{C})\ge n- \ord_n(2) w$
\item[(iv)] If $n=2^m-1$, $\delta _1= 2w+1$, and $\delta _1 < 2^{\lceil m/2\rceil}+3$, hence $rank(\mathcal{C})=2^m-1-mw$
\item[(v)] If $n=2^m-1$,  $\delta_1=2^h-1$, then $d_H(\mathcal{C})=\delta_1$
\item[(vi)] If $n=2^m-1$, then $d_H(\mathcal{C})\le 2\delta_1-1$
\item[(vii)]If $n= a\delta_1$, then $d_H(\mathcal{C})=\delta_1$
\item[(viii)] If $n=2^m-1$, $\delta_1=2w+1$, then if $2^{sw} < \sum_{i=0}^{w+1} \binom{n}{i}$, then $d_H=2w+1$
\end{itemize}
\end{thm}

\pf Part (i) follows from the $BCH$ like-bound for the Lee distance
of cyclic codes over $R$ given by~\cite[Theorem 7]{udaya2} and from
Part (ii) of Theorem~\ref{th:dis}. The other assertions follows from
Part (i), Theorem~\ref{th:dis} and the results for BCH codes over
fields in~\cite[Chap. 9]{macwilliams}. \qed
\begin{ex}
For $n=63, \delta_0=11$, and $\delta_1=9$ we have a $BCH(63, 11,9)$
code over $R$, with $2^{75}$ codewords, minimum Lee distance 11, and
minimum Hamming weight 9.
\end{ex}

\section{DNA Codes}

This section presents the design of DNA codes. First we give the
following definition.
\begin{defi}
\label{def2:dna} A code $\mathcal{C}$ is said to be reversible,
respectively complement, if it satisfies $x^r \in \mathcal{C}$ for
all $x\in \mathcal{C}$, respectively $x^c \in \mathcal{C}$ for all
$x \in \mathcal{C}$. A code $\mathcal{C}$ is said to be
reversible-complement if $x^{rc} \in \mathcal{C}$ for all $x\in
\mathcal{C}$. A reversible-complement cyclic code is a cyclic code
which is also reversible complement.
\end{defi}

\subsection{The Reverse-Constraint}

A sufficient condition for a code to satisfy the reverse constraint
is to be invariant under the reverse permutation $\sigma_R$ given by
$\sigma_R(c_0,c_1\ldots, c_{n-1})=(c_{n-1},\ldots c_1,c_0)$. If
$c(x)=c_0+c_1x+\ldots c_{n-1}x^{n-1}$ is a codeword of a cyclic
code, we have $\sigma_R(c(x))=x^{n-1}c(x^{-1})$. Codes invariant
under the action of $\sigma_R$ are called reversible.
\begin{defi}
For $f(x)\in R[x]$, let $f(x)^*=x^{\deg (f)} f(1/x)$ be the
reciprocal polynomial of $f(x)$. If equality holds between $f(x)$
and $f(x)^*$, we say that the polynomial is self-reciprocal.
\end{defi}
\begin{lem}
\label{lem:abua}(\cite[Lemma 4]{abualrub} Let $f(x)$ and $g(x)$ be
two polynomials in $R[x]$ with $\deg f(x) \ge \deg f(x)$. Then the
following holds.
\begin{itemize}
\item[(i)] $[f(x)g(x)]^*=f(x)^*g(x)^*$
\item[(ii)] $[f(x)+g(x)]^*=f(x)^*+x^{\deg f- \deg g }g(x)^*$
\end{itemize}
\end{lem}

The following result due to Massey~\cite[Theorem 1]{massey}
characterizes the reversible codes over fields.
 \begin{lem}
 \label{lem:massey1}
A cyclic code over a finite field $\FF_q$ generated by a monic
polynomial $g(x)$ is reversible if and only if $g(x)$ is
self-reciprocal.
 \end{lem}
A cyclic code $\mathcal{C}=\langle f_0|uf_1 \rangle$ over $R$ is
said to be free if it satisfies $\mathcal{C}=\langle f_0 \rangle$,
i.e., $f_1=f_0$. By a proof similar to that of
Lemma~\ref{lem:massey1} we have the following result.
\begin{lem}
Let $\mathcal{C}=\langle f(x) \rangle$ be a free cyclic code over
$R$ generated by a monic polynomial $f(x)|x^n-1$. Then $C$ is
reversible if and only if $f(x)$ is self-reciprocal.
\end{lem}
Conversely, if the code is not free the situation is different as we
prove in the following theorem.
\begin{thm}
\label{th:massey2} Let $\mathcal{C}=\langle f_0|uf_1 \rangle$ be a
cyclic code of odd length $n$. Then $\mathcal{C}$ is reversible if
and only if $f_0$ and $f_1$ are self-reciprocal.
\end{thm}
\pf We have a natural ring-morphism $\Psi :R \mapsto \FF_2$ defined
by $\Psi(a)=a^2\mod 2.$ Then $\Psi$ can be extended as follows
$\Phi:\mathcal{C} \mapsto \FF_2[x]/(x^n-1)$ defined by
\[\Phi(c_0+c_1x+\ldots+c_{n-1}x^{n-1})=\Psi(c_0)+\Psi(c_1)x+\ldots \Psi(c_{n-1})x^{n-1}\]
From \cite{abualrub}, we have the ideal $ker(\Phi)=\langle uf_1
\rangle$ and $\Phi(\mathcal{C})=\langle f_0 \rangle$. Note that the
last ideal is in $\FF_2[x]$. Since we have assumed that
$\mathcal{C}$ is reversible then $\Phi(\mathcal{C})=\langle f_0
\rangle$ is also reversible. Hence from Lemma~\ref{lem:massey1} the
polynomial $f_0$ is self-reciprocal. Since $f_1$ is a binary
polynomial that divides $f_0$, there exists a polynomial $g \in
\FF_2[X]$ such that $f_0=f_1g$. We have that
$f_0^*=(f_1g)^*=f_1^*g^*=f_0=f_1g$ since $f_1^*$ and $f_1$ are in
$\FF_2[x]$ with the same leading coefficient the same degree and the
same constant term, and the polynomial $f_0|x^n-1$ has simple roots,
so then $f_0=f_0^*$ and $g=g^*$.

Assume now that $f_0$ and $f_1$ are self-reciprocal, and let $c(x)$
be a codeword of $\mathcal{C}$. Then there exists $\alpha_0(x)$ and
$\alpha_1(x)$ in $R[x]$ such that
$c(x)=\alpha_0(x)f_0(x)+\alpha_1(x)uf_1(x)$. Using
Lemma~\ref{lem:abua} and the fact that $f_0(x)$ and $f_1(x)$ are
self-reciprocal,
$c(x)^*=\alpha_0^*(x)f_0(x)+\alpha_1^*(x)ux^mf_1(x)$, which means
that $c(x)^*$ is in $\mathcal{C}$. Since the code $\mathcal{C}$ is
cyclic, $x^{n-r-1}c^*(x)=x^{n-1}c(x^{-1}) \in \mathcal{C}$ means
that the reverse permutation leaves the code $\mathcal{C}$
invariant. Hence it is reversible. \qed

\subsection{The Reverse-Complement Constraint}

From Definition~\ref{def2:dna}, we have that a linear code which is
reversible complement satisfies the reverse-complement constraint.
\begin{lem}
\label{lem:1} If $\mathcal{C}$ is a reversible-complement cyclic
code, then $C$ contains the codeword
\[
u\mathbb{I}(x)=u+ux+\cdots +ux^{n-1}.
\]
\end{lem}
\pf Since $\mathcal{C}$ is linear, then $(0,\ldots,0)\in
\mathcal{C}$. Also, $\mathcal{C}$ is reversible complement, so that
$(0,\ldots,0)^{rc}=(u,\ldots,u)\in \mathcal{C}$. The last codeword
corresponds to the polynomial $u\mathbb{I}(x)= u+ux+\cdots
+ux^{n-1}$. \qed
\begin{thm}
\label{th:rec} Let $\mathcal{C} =\langle f_0+uf_1\rangle=\langle
f_0|uf_1\rangle$, be a cyclic code over $R$ of length odd $n$, with
$f_1 |f_0|x^n-1$ in $\F_2$. If $\mathcal{C}$ is a
reversible-complement code then we have $u\mathbb{I}(x)\in
\mathcal{C}$, $f_0(x)$ and $f_1(x)$ are self-reciprocal.
\end{thm}
\pf From~Lemma~\ref{lem:1} we have $u\mathbb{I}(x)\in \mathcal{C}$.
Now, let $f_0(x)=a_0+a_1x+\ldots a_r x^r$. Since $f_0\in \F_2[x]$
and $f_0|x^n-1$, then $f_0(x)=1+a_1x+\ldots+a_{r-1}x^{r-1}+x^r$. The
vector representation of $f_0(x)$ is equal to
$v=(1,a_1,\ldots,a_{r-1},1,0,\ldots,0)$. Hence
$v^{rc}=(\hat{0},\ldots,\hat{0},\hat{1},\hat{a}_{r-1}\ldots
\hat{a}_1,\hat{1}) \in \mathcal{C}$, and $f_0^{rc}(x)=u+ux+\ldots
+ux^{n-r-1}+\overline{u}x^{n-r}+\hat{a}_{r-1}x^{n-r+1}+\ldots
\hat{a}_{1}x^{n-1}\in \mathcal{C}$. Since $\mathcal{C}$ is linear,
we have $f_0^{rc}(x)+ u\mathbb{I}(x) \in \mathcal{C}$. Using
(\ref{eq:hat}) and the fact that the characteristic of $R$ is 2 we
obtain
\[
 f_0^{rc}(x)+u\mathbb{I}(x)=
x^{n-r}(1+a_{r-1}x+\ldots+a_1x^{r-1}+x^r) \in \mathcal{C}.
\]
Now multiplying $f^{rc}(x)+u\mathbb{I}(x)$ by $x^{r}$ and using the
fact that this operation is modulo $x^n-1$, we obtain
$f_0(x)^*=1+a_{r-1}x+\ldots+a_1x^{r-1}+x^r \in C$. Since
$\mathcal{C}=\langle f_0|uf_1\rangle$, there exists $k_0(x), k_1(x)
\in R[x]$ such that $f_0(x)^*=k_0(x)f_0+ uk_1(x)f_1(x)$. Multiplying
both sides of the previous equality by $u$ gives
\[
uf_0(x)^*=uk_0(x)f_0(x),
\]
but since $f_0(x)^*, f_0(x) \in \F_2[x]$ have the same degree, the
same leading coefficients and the same constant term, it must be
that $k_0(x)=1$. This means that $f_0(x)$ is self-reciprocal.

Now let $uf_1(x)=u(1+b_1x+\ldots+b_{s-1}x^{s-1}+x^s)$. Then
\[
uf_1(x)^{rc}=u+ux+ux^2+\ldots+ux^{n-s-2}+\hat{u}x^{n-s-1}+\hat{ub}_{s-1}x^{n-s-2}+\ldots
\hat{ub}_{1}x^{n-2}+\hat{u}x^{n-1}\in C
\]
and hence $uf_1(x)^{rc}+u\mathbb{I}(x)\in \mathcal{C}$. Using
(\ref{eq:hat}) and the fact that the characteristic of $R$ is 2 we
obtain that the last polynomial is equal to
$ux^{n-s-1}+ub_{s-1}x^{n-s}+\ldots+ub_1x^{n-2}+ux^{n-1}$. Hence
$uf_1^*\in C$, and for $f_0$ we obtain that $f_1(x)^*=f_1(x)$.
 \qed
Now we prove that the condition given by Theorem~\ref{th:rec} is
also sufficient.
\begin{thm}
\label{th:self3} Suppose $\mathcal{C}=\langle f_0| uf_1 \rangle$ is
a cyclic code of odd length $n$ over $R$ with $f_1|f_0|(x^n-1) \in
F_2[x]$. If $u+ux+\ldots+ux^{n-1}\in \mathcal{C}$ and $f_0,f_1$ are
self-reciprocal then $\mathcal{C}$ is a reversible-complement code.
\end{thm}
\pf Let $c(x)\in \mathcal{C}$. We must prove that $c(x)^{rc}\in
\mathcal{C}$. Since $\mathcal{C}=\langle f_0| uf_1 \rangle$, there
exist $ \alpha_0(x), \alpha_1(x)\in R[x]$ such that
\[
c(x)= \alpha_0(x)f_0(x)+ \alpha_1(x)uf_1(x).
\]
Taking the reciprocal and by repeated use of Lemma~\ref{lem:abua}
and the fact that $f_0(x)$ and $f_1(x)$ are self-reciprocal we have
\[
c(x)^*= \alpha_0(x)^*f_0(x)+ \alpha_1(x)^*ux^mf_1(x).
\]
This gives that $c^*(x)$ is in $\mathcal{C}$. Since $\mathcal{C}$ is
cyclic, $x^{n-t-1}c(x)=c_0x^{n-t-1}+c_1x^{n-t}+\ldots+c_t x^{n-1}\in
\mathcal{C}$. It was also assumed that $u+ux+\ldots ux^{n-1}\in
\mathcal{C}$, which leads to
\[
u+ux+\ldots ux^{n-1}+c_0x^{n-t-1}+c_1x^{n-t}+\ldots+c_tx^{n-1}\in
\mathcal{C}.
\]
This is equal to $u+ux+\ldots+\ldots
ux^{n-t-2}+(u+c_0)x^{n-t-1}+\ldots (u+c_t)x^{n-1}= u+ux+\ldots
ux^{n-t-2}+\hat{c_0}x^{n-t-1}+\ldots+ \hat{c_t}x^{n-1}$, which is
precisely $(c^*(x)^{{rc}})^*=c(x)^{rc}\in \mathcal{C}$. \qed

\begin{cor}
\label{cor:dual} Let $\mathcal{C}$ be a cyclic code with odd length
$n$. Then if $u+ux+\ldots +ux^{n-1} \in \mathcal{C}$ and if there
exists an $i$ such that
\begin{equation}
\label{eq:dual} 2^i\equiv -1 \mod n,
\end{equation}
then the code $\mathcal{C}$ is a reversible-complement code.
\end{cor}
\pf Let $\mathcal{C}=\langle f_0 | u f_1 \rangle$ be a cyclic code.
The polynomials $f_i$ are divisors of $x^n-1$ in $\F_2$. The
decomposition into the product of minimal polynomials is given by
$x^n-1=\prod M_i(x)$. Each $M_i$ corresponds to a cyclotomic class
$Cl(i)$. Equation~(\ref{eq:dual}) gives that $Cl(1)$ is reversible
and hence all the cyclotomic classes are reversible. Thus each
minimal polynomial is self-reciprocal, and from Lemma~\ref{lem:abua}
the polynomials $f_i$ are self-reciprocal. Then from
Theorem~\ref{th:self3} $\mathcal{C}$ is a reversible-complement
code. \qed
\begin{rem}
It is obvious that the Hamming distance constraint is satisfied for
a linear code. Furthermore, from Theorem~\ref{th:rec} a cyclic code
$\langle f_0|uf_1 \rangle$ is reversible-complement when $f_0$ and
$f_1$ are self-reciprocal. Hence from Theorem~\ref{th:massey2} the
code is reversible.
\end{rem}

 \subsection{BCH-DNA Codes}

Now the construction of BCH-DNA codes is considered.
\begin{thm}
\label{th:bchDNA} Let $\mathcal{C}=BCH(n,\delta_0,\delta_1)$ be a
BCH code over $R$ of length $2^m+1$ with $m\geq1$. then the code
$\mathcal{C}$ is a DNA code over $R$.
\end{thm}
\pf Since $\mathcal{C}$ is a cyclic code, the polynomial $\mbox{lcm}
(M_i)\, 1\leq i \leq \delta_1-1$ is a codeword of $\mathcal{C}$.
Hence the codeword $ \prod_{i=1}^s M_i=x^n-1/(x-1)=1+x+\ldots
x^{n-1}\in \mathcal{C}$. Furthermore, we have $2^m \equiv -1 \mod
n$. Then form Corollary~\ref{cor:dual} we obtain that $C$ is a DNA
code.\qed

\begin{ex}
We have the existence of a $BCH(65,11,9)$ code which is a DNA code
with $2^{34}$ codewords and Lee minimum distance equal to the
Hamming minimum distance of 13.
\end{ex}
More generally by a same proof as Theorem~\ref{th:bchDNA} we can
have a BCH-DNA code of length $n$ satisfying (\ref{eq:dual}).
\begin{ex}
The code $BCH(43,7,3)$ is a BCH-DNA code with $2^{72}$ codewords and
minimum Lee distance 6. The binary image by the Gray map gives an
optimal binary code $[86,72,6]$~\cite{grassl}.
\end{ex}

\section{The $GC-$ Weight}

As explained in the introduction, DNA codes with the same
$GC-$content in all codeword ensure that the codewords have similar
thermodynamic characteristics (e.g., melting temperature).

\begin{lem}
\label{lem:codew} Let $\mathcal{C}=\langle f_0|uf_1 \rangle$ be a
cyclic code over $R$. Then the the code $uTor(\mathcal{C})=\langle
uf_1\rangle $ is the subcode of $C$ containing all codewords of $C$
a multiple of $u$.
\end{lem}
\pf Let $\mathcal{C}_u$ be the subcode of $\mathcal{C}$ containing
all codewords with nonzero elements $u$. Then it is obvious that the
code $uTor(\mathcal{C})$ is a subset of $\mathcal{C}_u$. Let $c$ be
a codeword of $\mathcal{C}_u$, hence
$c=k_0(x)f_0(x)+uk_1(x)f_1(x)=ug(x)$ with $k_0,k_1,g \in R[x]$. The
codewords $ug(x)$ have coordinates $0$ or $u$ so that we may write
$ug(x)=uf(x)$, with $f(x)$ a binary polynomial. Since $f_1|f_0$, we
obtain $f_1|f$, and hence $\mathcal{C}_u=uTor(\mathcal{C})=\langle
uf_1\rangle$. \qed
 \begin{thm}
 \label{th:due}
The $GC-$weight of $\mathcal{C}=\langle f_0 |uf_1\rangle$ is given
by the Hamming weight enumerator of the binary cyclic code $\langle
f_1 \rangle$.
 \end{thm}
\pf

The $GC-$content is obtained by multiplying the codewords of
$\mathcal{C}$ by $u$, and from Lemma~\ref{lem:codew} we have
$\mathcal{C}_u= \langle uf_1\rangle$. Hence the $GC-$content is
given by the Hamming weight of the binary code generated by $f_1$.
\qed

\section{Infinite Families of DNA Codes with Fixed $GC-$content}

 \subsection{DNA Codes from the Simplex Codes}

The binary simplex code $S_m$ is a code with parameters
$[2^m-1,m,2^{m-1}]$ and all nonzero codewords of weight $2^{m-1}$.
This is the dual of the $[2^m-1,2^{m}-1-m,3]$ Hamming code (which is
also a BCH code of designed distance 3. Then $S_m$ is cyclic code
with generator polynomial $h^*(x)$, which is the reciprocal of the
parity check polynomial $h(x)=x^n-1/M_1(x)$. If $Cl(1)$ is a
reversible class, then $h^*(x)=h(x)$, and it is given by
$h^*(x)=\frac{x^n-1}{M_1^*(x)}$. The simplex code is optimal in the
sense of the constant $GC-$content property. It suffice to consider
the free cyclic code over $R$ generated by $h^*(x)$. This gives a
cyclic codes over $R$ with $4^m$ codewords and constant $GC-$weight
$2^{m-1}$. Note that this DNA code contains more codewords than the
code constructed from the binary simplex code given by the called
Construction B2~\cite{olgica}.
\begin{ex}
For $m=4$, respectively $m=5$, we have a cyclic code of length 15,
respectively 31, containing 256 codewords with the same $GC-$Content
equal to 8, respectively 1024 codewords with the same $GC-$content
equal to 16. Usually the $GC-$content is required to be in the range
$30\%- 50\%$ of the length of the code.
\end{ex}

\subsection{DNA Codes from the Zetterberg Codes}

A binary code $\mathcal{C}$ is said to be irreducible if it is the
dual of a cyclic binary code generated by a minimal polynomial
associated with a primitive $n$th root of unity $\alpha$. Let $m >0$
and $n=2^m+1$, then $\ord_n(2)=2m$. Let $\beta$ be a primitive
element of $F_{2^{2m}}$, so that $\alpha=\beta^{2^m-1}$ is a
primitive $nth$ root of unity with splitting field $\FF_{2^{2m}}$.
Then the minimal polynomial associated with $\alpha$ is denoted by
$M_1=\prod_{i\in Cl(1)}(x-\alpha^i)$, and $\deg M_1= \ord_n(2)=2m$.
The binary cyclic code $\mathcal{C}_z$ generated by $M_1$ is called
the Zetterberg code. It is easily determined that the weights of
$C_z$ are symmetric, since it is a binary code which contains the
all-one codeword $\mathbb{I}(x)$. The parameters of $\mathcal{C}_z$
are given by the following theorem.
\begin{thm}(\cite[Theorem 16]{ding},\cite[Theorem 5.4]{shoof})\\
If $m \equiv 1 (\mod 2)$, then $C_z$ has parameters
\[
[2^m+1,2^m+1-2m,3].
\]
$A_3=A_{2^m-2}=\frac{2^m+1}{3}$ and $A_4=A_{2^m-3}=0$.

If $m\equiv 0 \mod 2$, then $C_z$ has parameters
\[
[2^m+1,2^m+1-2m,5 \le d \le 6 ].
\]

The asymptotic behavior of $A_i$ is given by
\[
B_i=\frac{1}{2^{2m}}\binom{2^m+1}{i}+\mathcal{O}(2^i), 2^m
\rightarrow \infty.
\]
\end{thm}

The dual code $C_z^{\bot}$, is called the irreducible Zetterberg
code. It is a cyclic code generated by the polynomial $h^*(x)$,
where $h(x)=\frac{x^n-1}{M_1}$. Since $Cl(1)$ is a reversible class,
$M_1$ is self-reciprocal and hence $h^*=h$. This gives that the
dimension of $\mathcal{C}_z^{\bot}$ is equal to $2m$. We have the
following result
\begin{lem}\label{lem:edel}(\cite{edel})
All the weights of $C_z^{\bot}$ are even and the non zero-weight are
\[
a_{2i}=(2^m+1)m_i
\]
with $m_i$ a constant dependant on $i$. The (even) minimum distance
$d_z^{\bot}$ is bounded by $d_z^{\bot}> \frac{2^m+1}{2}-
\sqrt{2^m}$.
\end{lem}
\begin{prop}
\label{prop:zetter} The code $\mathcal{C}_0=\langle
\frac{x^n-1}{(x-1)M_1}\rangle$ has parameters
\[
[2^m+1,2m+1,d=\min(d_z^{\bot}, 2^m+1-d_z^{\bot})].
\]
The weight enumerator of $\mathcal{C}_0$ is
\[
\sum a_{2i}(x^{2i}+x^{2m+1-2i})
\]
where the $a_{2i}$ are the weights of the dual Zetterberg code given
by Lemma~\ref{lem:edel}.
\end{prop}
\pf The generator of $\mathcal{C}_0$ has degree $2^m-2m$, hence the
dimension is $2m+1$. The code $C_z^{\bot}$ is a subcode of
$\mathcal{C}_0$, and the all-one codeword $\mathbb{I}$ is in
$\mathcal{C}_0$. The weights of $\mathcal{C}_0$ are symmetric since
it is a binary code that contains $\mathbb{I}$. Let $c_{2i}$ be a
codeword of $C_z^{\bot}$ of weight $2i$. Then the codeword
$\mathbb{I}-c_{2i}$ is in $\mathcal{C}_0$ and has weight $2^m+1-2i$.
Hence there are at least $a_{2i}$ codewords in $\mathcal{C}_0$ of
weight $2i$ and at least $a_{2i}$ codewords with weight $2^m+1-2i$.
The total number of codewords in $\mathcal{C}_0$ is $2^{2^m+1}$,
whereas the total number of codeword in $\mathcal{C}_z^{\bot}$ is
$2^m$. Hence this gives that the weight enumerator of
$\mathcal{C}_0$ is $\sum a_{2i}\left(x^{2i}+x^{2m+1-2i}\right)$. The
minimum distance of $\mathcal{C}_0$ is given by the minimum of
$d_z^{\bot}$ and $2^m+1-d_z^{\bot}$.
\begin{thm}
Let $ \mathcal{C}=\langle f_1 \rangle$ be the free cyclic code of
$R$ generated by $f_1=\frac{x^{2^m+1}-1}{(x-1)M_1}$. Hence
$\mathcal{C}$ is a DNA code with $2^{2(2m+1)}$ codewords, minimum
distance $d=\min(d_z^{\bot}, 2^m+1-d_z^{\bot})$, and $GC-$weight
given by
\[
\sum a_{2i}(x^{2i}+x^{2m+1-2i})
\]
where the $a_{2i}$ are the weights of the dual Zetterberg code given
by Lemma~\ref{lem:edel}.
\end{thm}
\pf Since $f_1$ is self-reciprocal and the codeword $u \mathbb{I}$
is in $\mathcal{C}$, from Theorem~\ref{th:rec} the code is a DNA
code. From Theorem~\ref{th:due} the $GC-$content is given by the
weight distribution of $\mathcal{C}_0$, which is given by
Proposition~\ref{prop:zetter}. Hence the result. \qed

\subsection{DNA Codes from the Reed--Muller Codes}

From Theorem~\ref{th:bchDNA}, there exist BCH-DNA codes of length
$2^m+1$. In the following section, we consider the construction of
families of DNA codes of length $2^m-1$ with fixed $GC-$content. We
begin by proving the following result.
\begin{prop}
\label{thm:-1splitting} Let $n$ be an odd integer. Then if $\ord_n
(2)$ is even there exists a $2$cyclotomic class modulo $n$ which is
reversible.
\end{prop}
\begin{enumerate}
\item If $n=p$ is a prime, we assume that $\ord_n (2)=2w$ is even $2^{2w}\equiv 1\bmod n$.
Hence $n|(2^{w}-1)(2^{w}+1)$. Since $n$ is prime and cannot divide
$2^{w}-1$ (because of the order), we have $2^{w}=-1 \bmod n$ which
gives that $Cl(1)$ is reversible.
\item If $n =p^{\alpha}$, we first have to prove the following implication
\[
\ord_{p^{\alpha}}(2) \text{ is even }\Rightarrow\ord_p(2) \text{ is
even}.
\]
Assume $\ord_{p^{\alpha}}(2)\text{ even}$ and $\ord_p(2) \text{
odd}$. Then there exist $i>0$ odd such that $2^i\equiv 1 \mod p
\Leftrightarrow 2^i=1+kp$. Hence $2^{ip^{\alpha
-1}}=(1+kp)^{p^{\alpha- 1}} \equiv 1\mod p^{\alpha}$, because $(1+
kp)^{p^{\alpha -1}} \equiv (1+kp^{\alpha}\mod p^{\alpha+1})$ (the
proof of the last equality can be found in
\cite[Lemma~3.30]{demazure}). Hence
\begin{eqnarray}
\label{fermat} 2^ {ip^{\alpha -1}} \equiv 1\mod p^{\alpha}.
\end{eqnarray}
With $i$ odd and ${p^{\alpha -1}}$ odd, $\ord_{p^{\alpha}}(2)$ is
odd (because $\ord_{p^{\alpha}}(2) | i p^{{\alpha}-1}$), which is
absurd. Hence $\ord_p(2)$ is even, then there exists some integer
$j$ such that $0<j< \ord_p (2)$, and $2^j\equiv -1\mod p$. Then from
(\ref{fermat}), we have $2^{jp^{\alpha -1}}\equiv -1\mod
p^{\alpha}$. This gives that $Cl(1)$ is reversible.
\item If $n = p_1p_2$ with $(p_1,p_2)=1$, since
$\ord_n(2)=\mbox{lcm}(\ord_{p_1}(2),\ord_{p_2}(2))$ is even, then
either $\ord_{p_1}(2)$ or $\ord_{p_2}(2)$ must be even. Assume that
$\ord_{p_1}(2)$ is even. Then there exits $1\leq k\leq \ord_{p_1}(2)
\textrm{ such that } q^k\equiv -1\mod  p_1$. Therefore
$q^k(n-p_2)\equiv -(n-p_2)\mod n,$ with $k\leq \ord_{p_1}(2)$.
\item If $n=p_1^{\alpha_ 1}p_2^{\alpha_ 2}$ with $(p_1,p_2)=1$, we know that $\ord _n (2) =\mbox{lcm}(\ord_{{p_1}^{\alpha_1}}(2),\ord_{{p_2}^{\alpha_2}}(2))$.
Then if $\ord_{p_1^{\alpha_ 1}p_2^{\alpha_ 2}}(2) $is even we have
either $\ord_{p_1}(2)$ or $\ord_{p_2}(2)$ is even. Therefore it
suffices to repeat the process in case 3 above.

Hence the generalization to any $n$ such that $\ord_n(2)$ is even.
\end{enumerate}
\qed

Now we consider the family of second order Reed-Muller
codes~\cite[Ch. 13-15]{macwilliams}. The punctured second order
Reed-Muller code $R^*(2,m)$ is a cyclic code of length $2^m-1$,
dimension $1+m+\frac{(m-1)m}{2}$, and generator polynomial
$g(x)=\prod_{1\le w_2(s) \le m-3} M_{s}, 1\le s \le 2^m-2$.
$R^*(2,m)$ contains the all one codeword and has minimum Hamming
distance $2^{m-2}-1$. The code $R^*(2,m)$ is a subset of the binary
BCH code $BCH_2(2^m-1,2^{m-2}-1)$ of designed distance $2^{m-2}-1$
and dimension $2^m-1-m(2^{m-3}-1)$. The binary weight distribution
of $R(2,m)$ is given in~\cite[p. 443]{macwilliams}. Since the codes
$R(2,m)$ are affine-invariant, we can apply~\cite[Theorem 14, Ch.
8]{macwilliams} to determine the weight distribution of the
punctured code $R(2,m)^*$. Since this is a well known infinite class
of codes with known weight distribution, it will be used to
construct DNA codes with the reverse-complement constraint and also
good $GC-$content.

Let $n=2^m-1$ be a positive integer. If $m$ is even then from
Proposition~\ref{thm:-1splitting} there exists at least one
reversible class modulo $n$. Let $g(x)\in \FF_2[x]$ be a monic
divisor of $x^n-1$ which generates the code $RM(2,m)^*$. This can be
decomposed as $g(x)= g_1(x)g_2(x)$ such that $g_1(x)$ is the product
of all non self-reciprocal minimal polynomials that divide $g(x)$,
and $g_2(x)$ is the product of all self-reciprocal minimal
polynomials that divide $g(x)$. Hence $g_2(x)$ is a self-reciprocal
polynomial, and the all one codeword is contained in the code
generated by $g_2(x)$. From Theorem~\ref{th:self3} we then have a
DNA code $\mathcal{C} = \langle g_2(x)\rangle$. This code contains
at least $A_i$ codewords with $GC-$content equal to $i$ where the
$A_i$ are the coefficients of the weight enumerator of $RM^*(2,m)$.
\begin{ex}
If $m=4,$ then $n=15$ and there are 5 cyclotomic classes. The only
reversible class is $Cl(5)$, and the generator of $RM(2,4)*$ is the
minimal polynomial associated with $Cl(1)$. Thus we cannot apply the
procedure above.

If $m=6,$ then $n=63$, $Cl(7)$ and $Cl(21)$ are reversible classes.
Furthermore $M_7M_{21} |g(x)$ the generator polynomial of
$RM(2,6)^*$. Hence $\langle M_7M_{21}(x) \rangle $ is a DNA code
over $R$ since it is generated by a self-reciprocal polynomial and
contains the codeword $u\mathbb{I}$. $RM(2,6)^*$ is a subcode of
$\langle M_7M_{21}(x) \rangle $. For a given weight $i$, this code
contains at least $A_i$ codewords of weight $i$ where the $A_i$ are
the coefficients of the weight polynomial of $RM(2,6)^*$. These
weights are given in the following table.
\begin{table}[h]
\label{table:2}
\begin{center}
\begin{tabular}{|c|c|}
\hline
 $i$&$A_i$ \\
\hline
 47or 15 &2604   \\
   23 or 39 &291648\\
27 or 35&888832\\
31& 3011220 \\
 \hline
\end{tabular}
\end{center}
\caption{The minimum number of codewords of weight $i$ in the DNA
code $\langle M_7(x) M_{21}(x) \rangle $}
\end{table}

\end{ex}

\end{document}